\documentclass[12pt,manuscript]{aastex}
\usepackage{lscape,amssymb}

\begin{document}

\shorttitle{Stellar Tidal Streams in the Local Volume}
\shortauthors{Mart\'{i}nez-Delgado et al.}

\title{Stellar Tidal Streams in Spiral Galaxies of the Local Volume: \\
A Pilot
  Survey with Modest Aperture Telescopes}

\author{David Mart\'{i}nez-Delgado\altaffilmark{1,2}, R. Jay
  Gabany\altaffilmark{3}, Ken Crawford\altaffilmark{4}, Stefano
  Zibetti\altaffilmark{1}, \\
 Steven R. Majewski\altaffilmark{5},
Hans-Walter Rix\altaffilmark{1}, J\"urgen Fliri\altaffilmark{2,6},
  Julio A. Carballo-Bello\altaffilmark{2},\\ Daniella C. Bardalez-Gagliuffi\altaffilmark{2,7}, Jorge Pe\~narrubia\altaffilmark{8},
  Taylor S. Chonis\altaffilmark{9}, \\
  Barry Madore\altaffilmark{10}, Ignacio
  Trujillo\altaffilmark{2}, Mischa Schirmer\altaffilmark{11}, David A. McDavid\altaffilmark{5}}

\altaffiltext{1} {Max Planck Institut f\"ur Astronomie, Heidelberg, Germany}
\altaffiltext{2} {Instituto de Astrof\'isica de Canarias, La Laguna, Spain}
\altaffiltext{3} {Black Bird Observatory, Mayhill, New Mexico, USA}
\altaffiltext{4} {Rancho del Sol Observatory, Modesto, California, USA}
\altaffiltext{5} {Department of Astronomy, University of Virginia, USA}
\altaffiltext{6} {GEPI, Observatoire de Paris, Meudon Cedex, France}
\altaffiltext{7} {Massachussetts Institute of Technology, Cambridge, MA, USA}
\altaffiltext{8} {Institute of Astronomy, University of Cambridge, UK}
\altaffiltext{9} {Department of Astronomy, University of Texas, Austin, USA}
\altaffiltext{10} {The Observatories of the Carnegie Institution of Washington, Pasadena, USA}
\altaffiltext{11} {Argelander Institut f\"ur Astronomie, Universit\"at Bonn, Germany}

\begin{abstract}

Within the hierarchical framework for galaxy formation, minor
 merging and tidal interactions are expected to shape all large galaxies
 to the present day. As a consequence, most seemingly normal disk galaxies
should be surrounded by spatially extended stellar 'tidal features' of low
surface brightness.  As part of a pilot survey for such interaction
signatures,
we have carried out ultra deep, wide field imaging of 8 isolated spiral
galaxies in  the Local Volume,  with data taken at small ($D=$0.1-0.5m)
robotic telescopes that provide exquisite surface brightness sensitivity
($\mu_{lim}(V)\sim 28.5$ mag/arcsec$^2$).

 This initial observational effort has led to the discovery of six previously
undetected extensive (to $\sim 30$kpc )  stellar structures in the
halos surrounding these galaxies, likely debris from tidally disrupted
satellites. In addition, we confirm and clarify several enormous stellar over-densities previously reported in the literature, but never before interpreted
  as tidal streams.

 Even this pilot sample of galaxies exhibits strikingly diverse
 morphological characteristics of these extended stellar features:
{\it great circle}-like features that resemble
  the Sagittarius stream surrounding the Milky Way,  remote
  shells and giant clouds of presumed tidal debris far beyond the
main stelar body, as well as jet-like  features emerging from
galactic disks. Together with presumed  remains of already disrupted
companions, our observations also capture surviving satellites caught
in the act of tidal  disruption.

A qualitative comparison with available simulations set in a
  $\Lambda$Cold Dark Matter cosmology (that model the stellar
halo as the result  of satellite disruption evolution) shows that
the extraordinary variety of stellar morphologies detected in
this pilot survey matches that seen in those simulations.
The common existence of these tidal features around
'normal' disk galaxies and the morphological match
to the simulations constitutes new evidence that these theoretical
 models also apply to a large number of other Milky Way-mass
disk galaxies in the Local Volume.

\end{abstract}


\section{INTRODUCTION}\label{int}

Galactic mergers have long been recognized as crucial agent in shaping
and evolving galaxies (Toomre \& Toomre 1972). Within the hierarchical galaxy
formation framework (e.g., White \& Frenk, 1991), dark matter halo
mergers are a dominant evolutionary driver on the scale of galaxies.
For all mass scales lower  than entire galaxy clusters, the merger of two DM
halos is followed quickly by the  merger of the (stellar) galaxies that
had been sitting at the halo's centers (e.g.,  Kauffmann et al. 1993).
 The most spectacular
manifestations of this process may be {\it major mergers} (i.e.,the
coalescence of galaxies with comparable mass), that usually
entails the destruction of any pre-existing stellar disk and may lead
to star-bursts. Such events have been relatively rare at least since
$z\sim 1$, with only a few percent of luminous galaxies being involved in an
ongoing major merger at any point in time (e.g., Robaina et al 2009).

However, {\it minor mergers} (i.e., the coalescence of a satellite
galaxy and its halo with a much more luminous and massive companion)
are expected to be significantly more common (e.g., Cole et al.
2000). Indeed, such minor mergers should remain frequent to the
present epoch in a $\Lambda$CDM cosmogony. As minor mergers do not
destroy pre-existing stellar disks (e.g., Robertson et al. 2006), signs
of recent or on-going minor mergers should be apparent around spirals,
the most common type of large galaxy. If the satellite galaxies become
tidally disrupted while still in an orbit that extends beyond the
stellar body of the larger galaxy companion, then they should form
stellar tidal 'features', which extend into the halo of the central
galaxy.

The observational consequences of this scenario, where the stellar halo of spiral
galaxies is essentially comprised of tidal stellar debris from merged
satellite galaxies, has been explored by Bullock \& Johnston (2005,
BJ05) and others (e.g., Tumlinson et al. 2009; Cooper et al.
2010). Satellites that merged on compact orbits or a long time ago have
phase-mixed into a seemingly smooth component by now. In contrast,
merger remnants that are only a few dynamical periods old, either
because they occurred recently or on orbits with $t_{orbit}\gtrsim 1$~
Gyr, should leave stellar streams, rings or plumes as the 'fossil
record' of their interactions. BJ05 showed that such tidal debris can
exhibit a wide range of morphologies and that such distinctive structural features
should be common, perhaps ubiquitous 
around normal disk galaxies. 
They also
showed that most of the features occur at very low surface brightness
($\mu_{V}\gtrsim 28.5$ mag/arcsec$^2$) and would therefore not be
recognizable in traditional images of nearby galaxies.

The Milky Way and the Andromeda galaxy, both resolvable into individual
stars so that low surface brightness streams can
more readily be seen,
show a wealth of (sub-)structure in the stellar distribution of their
outskirts.  The most spectacular cases are  the
Sagittarius tidal stream surrounding the Milky Way (e.g., Majewski et al. 2003)
and the Great Southern stream around the Andromeda galaxy
(e.g., Ibata et al. 2001), which have become archetype fossil records of
satellite galaxy mergers. But overall, the stellar halo structure
of both galaxies is complex (e.g., Majewski, Munn \& Hawley 1994; Belokurov et al 2006; Bell et al 2008;
McConnachie et al 2009).

Both simulations and empirical evidence suggest that there is a great
deal of galaxy-to-galaxy variation in the level and the epoch of
merging and hence variation in the amount and morphology of tidal debris. 
Therefore, a more than qualitative comparison between the predicted and
observed prevalence of stellar debris
around disk galaxies requires a much larger sample
that necessarily must include galaxies well beyond the Local Group.  The current models
predict that a survey of between 50 and 100 parent galaxies reaching
to a surface brightness of $\sim 30 mag/arcsec^2$ should reveal many tens of
tidal features, perhaps nearly one detectable stream per galaxy
(Bullock \& Johnston 2005; Johnston et al. 2008; Cooper et al. 2010).

However, at present the evidence for tidal streams beyond the Local
Group is mostly anecdotal, rather than systematic. The first cases of
candidate extragalactic tidal stream candidates were reported a decade ago by
Malin \& Hardlin (1997). Using special contrast enhancement techniques
on deep photographic plates, these authors were able to highlight two
possible tidal streams surrounding the galaxies M83 and M104. Subsequently,
deep CCD images of the nearby, edge-on galaxy NGC 5907 by Shang et
al. (1998) revealed an elliptically-shaped loop in the halo of this
galaxy. This was the most compelling example of an external tidal
stream up to now.  More recently, very deep images have clearly
revealed large scale, complex structures of arcing loops in the halos
of several nearby, NGC galaxies (NGC 5907: Martinez-Delgado et al. 2008;
NGC 4013: Martinez-Delgado et al. 2009; NGC 891: Mouhcine, Ibata \&
Rejkuba 2010) as well as more distant, anonymous galaxies (Forbes et
al. 2003).

These results suggest that a more systematic survey for tidal streams
in the nearby universe is not only practical but required as a new
way to constrain models of galaxy formation.  During the past few
years, we have initiated a pilot survey of stellar tidal streams in a
select number of spiral systems using modest telescopes operating at
very dark sites. Ultimately, the most basic question we will seek to
answer concerns the frequency of stellar streams in the Local
Volume. Our aim is to test theoretical predictions by comparing
substructure counts from our galaxy sample to cosmological
simulations.  But the models also make predictions about a number of
direct observational characteristics (such as the colors,
morphologies, spatial coherence and extent of halo substructures) that
can be tested with the results of our survey.

This paper describes the initial results of our pilot study on eight
nearby spiral galaxies. These systems were selected for study because
they were already suspected of being surrounded by diffuse-light
over-densities based on data collected from available surveys (e.g.,
POSS-II; SDSS-I) and previously published deep images posted on the
internet by amateur astronomers. While based on a biased sample of
systems pre-selected for substructures, our pilot study serves as a
proof of concept for the intended, more systematic survey of halo
substructure around spiral galaxies. It also  enabled us to resolve the
required observing strategies and data reduction methodologies.  The
results presented here come from a productive collaboration between
amateur and professional astronomers, dedicated to exploiting the
scientific potential of modest aperture telescopes.

\section{OBSERVATIONS}\label{discussion}

This pilot survey was conducted with three privately-owned
observatories equipped with modest-sized telescopes located in the USA
and Australia (see Table 1). 
Each observing site features very dark,
clear skies with seeing that is routinely below 1.5 arcseconds.  These
telescopes are manufactured by {\it RC Optical Systems} and follow a
classic Ritchey-Chretien design. The observatories are commanded with
on-site control computers that allow remote operation and control from
any global location with high-bandwidht 
web access. Each observatory uses
proven, widely available, remote desktop control software.  Robotic
orchestration of all observatory and instrument functions, including
multiple target acquisition and data runs, is performed using
available scripting software. 
We also make use of a wide
field instrument for those galaxies with an 
especially 
extended angular size
(e.g., NGC 5055; see Fig. 1).  For this purpose, we have selected the
Astro-Physics Starfire (APS) 160EDF6, a short focal length (f/7), 16
cm-aperture refractor that provides a FOV of $\sim$ 73.7 $\times$
110.6 arcmin (see Table 1).

Each telescope is equipped with a commercially available CCD camera.
The primary survey camera of the Black Bird Observatory (BBO) is the
SBIG STL-11000, which uses a Kodak KAI-11000M imaging sensor.  This
sensor consists of a 4008 $\times$ 2672 pixel array with 9 $\times$ 9
micron pixels. Some of the other facilities use a 16-megapixel camera
manufactured by Apogee Instruments, of Roseville (California).  The
Rancho del Sol (RdS) observatory uses the Alta-KAFO9000 imaging sensor
with 3056 $\times$ 3056 pixels with 12 $\times$ 12 micron . The
Moorook (MrK) observatory uses a similar CCD camera, the Apogee
Alta-KAF16803 with 4096 $\times$ 4096 pixels and a smaller pixel size
(9 $\times$ 9 micron).

The tidal stream detection strategy and the procedures used for data
acquisition and reduction of the data are the same as those
described in our previous papers from this project (Martinez-Delgado
et al. 2008; 2009). In summary, this strategy strives for multiple
deep exposures of each target using a wide bandpass, high throughput,
clear optical 
filter with a near-IR cut-off, also known as a {\it luminance
  filter, L}.  The typical cumulative exposure times are in the range
of 6 to 11 hours. Data reduction followed standard techniques
described in the afore-mentioned papers. The list of targets, together
with the telescopes and total exposure time used in each case are
given in Table 2. 

Photometric calibration of the luminance filter ($L$) images is not currently 
available.  So, to assess their depth and typical quality in terms of 
background and flat-fielding, we relied on 
images of six of our galaxies --- NGC1055, NGC1084, 
NGC3521, NGC4216, NGC4651 and NGC5866 --- obtained by 
the Sloan Digital Sky Survey ( SDSS, York et al. 2000; Data Release 7, 
Abazajian et al. 2009).  

Based on SDSS photometry, we 
derived photometric equations to convert the $L$-band counts into $g$-band 
magnitudes.  SDSS image mosaics were constructed as described in 
Zibetti, Charlot \& Rix (2009) and high $S/N$, $g - L$ color maps of these 
galaxies were obtained with {\sc adaptsmooth} (Zibetti 2009). Using these 
maps, we estimated the median zero point and the amplitude of the color terms,
which turns out to be of the order of 0.1 mag, at most.

 There are two main limitations to the depth that can be reached in imaging
low-surface brightness features: (i) photon 
noise and (ii) background fluctuations due to flat-field residual, 
internal reflections, ghosts, scattered light, etc.  We estimate the 
photon noise limit as the surface brightness corresponding to 5 times  
r.m.s. in  $2^{\prime\prime}$-diameter random apertures. 
For background fluctuations, we estimated the median sky level r.m.s. in
selected boxes, several tens to hundred arcseconds per side, spread around the galaxies. We find that the typical $2^{\prime\prime}$-diameter detection limit is $27.2\pm 0.2~\mathrm{mag}_g~\mathrm{arcsec}^{-2}$, while the typical background fluctuations correspond to $28.5\pm 0.5~
 \mathrm{mag}_g~\mathrm{arcsec}^{-2}$.\footnote{$g$-band AB
    magnitudes are typically 0.3 mag higher than V-band Vega
    magnitudes (see Smith et al. 2002).}  It is worth noting that for the
corresponding SDSS g-band images we 
measured 25 and 28.7, respectively. This shows that
our images are roughly 10 times deeper than the SDSS data in terms of 
photon statistics and are mainly limited by systematic 
background uncertainties, which are comparable to those of the SDSS data.  
This implies that our images have high efficiency in detecting sharp
or localized features but background fluctuations hampers our ability
to accurately measure smooth diffuse light.

\section{THE RANGE OF TIDAL STREAM MORPHOLOGIES IN THE LOCAL
  UNIVERSE}\label{discussion}

Figure 1 presents recent images of eight nearby spiral galaxies that were found
to be surrounded by very low surface brightness structures. In the
cases shown, such features were suspected to be present based on
various imaging sources including the SDSS-I and POSS-II. Our new
observations have enabled the discovery of six morphologically unique
tidal stream candidates and reveal the likely tidal nature of numerous
faint galaxy structures reported over the last forty years from
inspection of photographic plates. Interestingly, these features do
not have HI counterparts in the available radio surveys (e.g., THINGS:
Walter et al. 2008), nor do they display any evidence of recent star
formation.  This is consistent with features that are predominantly
stellar streams tidally stripped from dwarf satellite galaxies. The
reality of these enormous structures was confirmed through independent
images obtained with multiple telescopes given in Table 1; 
we present
the deepest of those images in Figure 1, with the particulars of these
images given in Table 2.  A detailed investigation of these streams
and their properties will be given in a forthcoming contribution
(Martinez-Delgado et al. 2010, in preparation).

Results of our first forays into low surface brightness imaging
included the detection of
gigantic looping structures, analogous to the Milky Way's Sagittarius
stream, around the spiral galaxy NGC 5907 (Martinez-Delgado et al. 2008).
Our image of NGC 5055 here
(Messier 63: Fig.1a) uncovers a similar, faint, arcing feature in its halo. This
presumed tidal stream was first detected in very deep images obtained
with the 16-cm APS apocromatic refractor. This galaxy's
giant surrounding structure extends $\sim$30 kpc from the center of
NGC 5055 and appears to be unrelated to any faint outer regions of the
extended disk that were previously detected by the GALEX space
telescope (and also, incidentally, detected in our data). Some
insights about this arc were previously reported by van der Kruit
(1978) using photographic plates. The position of the stream's
progenitor remains unknown --- it may be completely disrupted or may
lie hidden behind the galaxy's disk.

Our BBO image of the nearby galaxy NGC 1084 (Fig.1b) also displays three
giant disconnected plumes of similar width extending a large
galactocentric distance ($\sim$ 30 kpc) into its halo. Two of these
tails emerge in opposite directions from the inner region of the
galaxy while a third one appears completely disconnected from the galaxy.  
These
features were first detected after a close visual inspection of SDSS
images. However, it remains difficult to assert if this collection of
arcing features is associated with one or several different merger
events.

In addition to the remains of presumably long disrupted companions,
our data also capture the ongoing tidal disruption of satellite
galaxies that are still visible, seen as long tails extending from the
progenitor satellite. Perhaps, the most conspicuous examples can be
seen in the image of NGC 4216 (Fig. 1c).  This panoramic view of the
galaxy shows two satellites with distinct cores and extremely long tails that extend several
kiloparsecs into the principal galaxy's halo. The host galaxy also
displays a prominent thick disk with several pillars arising from
it. The nature of these features (tidal debris or ram pressure
signatures) is discussed at length in Martinez-Delgado et al. (2010).

Among the most conspicuous features found in our survey are coherent
structures that resemble an open umbrella and extending tens of
kiloparsecs into the host spiral's halo. These spectacular formations
are often located on both sides of the principal galaxy and display
long narrow shafts that terminate in a giant, partial shell of
debris. The most remarkable example so far detected is in NGC 4651, shown
in Figure 1d. This is also the brightest tidal stream detected in our
pilot survey (visible even in very short exposure times). The jet-like
feature is strikingly coherent and narrow.  This feature was
previously reported by Vorontsov-Velyaminov (1959) but never
interpreted as a stellar tidal stream. Moreover, our deep image shows
an additional, spectacular, crescent-shaped shell surrounding the east side of
the galaxy that should 
correspond to
the apocenter of the dwarf galaxy 
progenitor. Interestingly, a possible second arc on the western side of
the galaxy can also be seen in this image. This structure is less
obvious because it is partially hidden by the galaxy's disk.  This
suggests we are observing a moderately inclined structure projected
into the halo of NGC 4651.

The giant, diffuse, cloud-like structure seen in the halo of NGC 7531
(Fig.~1e) was first reported by Buta (1987) from an inspection of
photographic plates. This author classified it as a possible dwarf
companion (named A2311.8-4353) in orbit about this spiral galaxy.  The report
also offered some insights about tidal disruption. Our deeper images
clearly reveal the actual shape and extension of this intriguing
feature including the presence of small scale substructuring. Its size
is comparable to the disk of the host galaxy.  We suspect it might be
the shell component of an umbrella-like structure, like that discussed
above.

Our images have also reveal the presence of smooth large-scale
structures that contain coherent substructures spread above and below
the plane of several galaxies.  These examples (Fig. 1g and 1h) are
excellent candidates for systems with ``mixed-type" stream ensembles 
(see Sec. 4), although alternate origins cannot be
rejected. Perhaps the most remarkable system is NGC 1055 (Fig.~1h).
This galaxy displays a clear boxy-shaped inner halo sprinkled with a
plethora of coherent ``spikes'' that seem to emerge from the galaxy's
disk.  
But the most striking examples of spike-like structures that have
been detected in our survey thus far are those associated with NGC
5866 (Fig.~1g).  In this system, a collimated structure extends $\sim$
8 kpc into the galactic halo. This galaxy also displays more
substructure in its ``smooth'' component, such as a wedge of material
in the plane of the disk on its eastern side.
The morphology of this disk extension is consistent with the remnants
of a recent merger event (e.g., see model F in Fig.~6 in Cooper et
al. 2010). In a similar way, the diffuse structure detected around NGC
3521 (Fig.~1f) also contains some discernible substructure, such as an
almost spherical cloud of debris visible on its eastern side
and a large, more elongated cloud on its western side. Both structures
may represent debris shells belonging to an umbrella like structure, as
seen in the image of NGC 4651 (Fig.~1d), but their fuzzier 
appearance
could suggest that they were accreted much farther in the past.

\section{DISCUSSION AND FUTURE WORK}\label{discussion}

Our pilot survey of tidal streams associated with nearby galaxies has
revealed that many spiral galaxies in the Local Universe contain
significant numbers of gigantic stellar structures that resemble the
features expected from hierarchical formation.  Although we have only
explored a handful of galaxies, our collection already presents a wide
spectrum of morphologies for these stellar features.  Some of them
maybe have analogs in the Milky Way --- e.g., (i) great arc-like features
(labeled {\it A} in Fig.~2) that resemble the Milky Way's Sagittarius,
Orphan and Anticenter streams (e.g., Majewski et al. 2003, Belokurov
et al. 2006, 2007b, Grillmair 2006) and (ii)  enormous clouds of debris that
 resemble our current understandings of
the expansive Tri-And and Virgo overdensities and the Hercules-Aquila
cloud in the Galactic halo (Rocha-Pinto et al. 2004, Belokurov et al. 2007a,
 Martinez-Delgado et al. 2007, Juric et al. 2008). Our
observations also uncover enormous features resembling giant
``umbrellas'' (labeled {\it U} in Fig.~2), isolated shells, giant
plumes of debris (labeled {\it GP} in Fig.~2), spike-like patterns (labeled
{\it S} in Fig.~2) emerging from galactic disks, long, tighly
coherent streams with a central remnant core (labeled {\it PD} in Fig.~2) and
large-scale diffuse forms that are possibly related to the remnants of
ancient, fully disrupted satellites.

Remarkably, the diverse morphologies of stellar tidal features
detected in our pilot data nearly span the range of morphologies seen
in cosmologically motivated simulations. Therefore, they
already represent the most comprehensive evidence matching and
supporting the detailed hierarchical formation scenario predictions
for galaxies similar to the Milky Way. We illustrate this through
comparison with the set of eleven available snapshots featuring stellar halo
models from BJ05. Each model was constructed
with different merger histories in a $\Lambda$CDM Universe and
provides an external, panoramic view of surviving tidal debris (see
Fig.2) from about one hundred satellites. These low-mass systems were
``injected'' into a central halo potential along orbits whose
distributions is consistent with current cosmological models.

Even when limiting the output of the simulations to a surface brightness
comparable to our observational limit, the different stream
morphologies seen fossilized in these nearby spiral halos (see Fig.~1)
can be easily identified in snapshots of the model halos as well. This
is illustrated by Figure 2, which compares the most conspicuous types
of tidal debris detected in our survey with those visible in the model
snapshots for three different assembly histories.\footnote{This
 tight correspondence of observation and theory also holds for the comparison 
 with the more recent simulations by Cooper et al. (2010)."} 
From an analysis of their models, Johnston et
al. (2008: see their Fig. 1) concluded that the observed stream
morphology is principally dependent upon the progenitor satellite's
orbit and accretion epoch.  For example, {\it great circle} features
(like those seen around NGC 5907 and M63) apparently arise from
near circular orbit accretion events that occurred 6-10 Gigayears ago.
Straight narrow features with associated shells (e.g., the spikes in
NGC 5866 and the umbrella shaped structure in NGC 4651, labeled ``Sp"
and ``U" respectively in Fig.~2) were formed in a similar epoch from
low-mass satellites in almost radial orbits. Finally, the large-scale
diffuse structures observed around NGC 1055 and NGC 5866 could
correspond to the mixed-type category pointed out by these authors, in
which case they represent the debris of one accreted satellite that
occurred longer than 10 Gigayears ago that have had time to fully mix
along its orbit.

Figure 2 also illustrates how the stochastic nature of halo formation
(Cooper et al. 2010) leads to a large variety of substructures in the
outer regions of the galaxies: each halo displays a unique and very
complex pattern of stellar debris caused by different defunct
companions. This morphological variance among different galaxies
should be largest for brighter streams, typically formed by massive
and quite recent mergers. This would explain why our modest sample,
which was constructed to contain comparatively prominent streams in
the Local Universe, reveals such a wide variety of detected
streams. Despite the biased sample selection, the results presented
here may constitute the first comprehensive observational evidence
to support that the predicted great diversity of stellar halos/stream
morphologies is 
actually present in Nature.

Encouraged by this pilot survey, we have embarked on the first
systematic search for stellar tidal streams in a complete,
volume limited sample of spiral galaxies up to 15 Mpc (i.e., the Local
Volume). This will result in the first comprehensive census 
of stellar stream structures within the Local Volume and it
 will enable meaningful statistical comparisons with cosmological
simulations. The frequency of streams, their stellar populations and
their morphologies will help reveal the nature of the progenitors and
lend insights into the underlying structure and gravitational
potential of the massive dark matter halos in which they reside. This
will thereby offer a unique opportunity to study the apparently
still dramatic last stages of galactic assembly in the Local
Universe. In this regard, the survey will be complementary to (and
directly inform the interpretation of) local galactic `archaeological'
data from 
resolved galaxies like
M31 and the Milky Way.

We thank K.V. Johnston for providing the models used in this paper and
for useful discussion. We also thank G. van de Venn, David
Valls-Gabaud, Andrew Cooper and M. A. Gomez-Flechoso for useful
discussions.  D. M-D acknowledges funding from the Spanish Ministry of
Education (Ramon y Cajal fellowship and research project AYA 2007-65090)
and the Instituto de Astrofisica de Canarias (proyect 3I. SRM appreciates support from NSF grant AST-0807945.

\newpage

\begin{figure}
\epsscale{.70}
\plotone{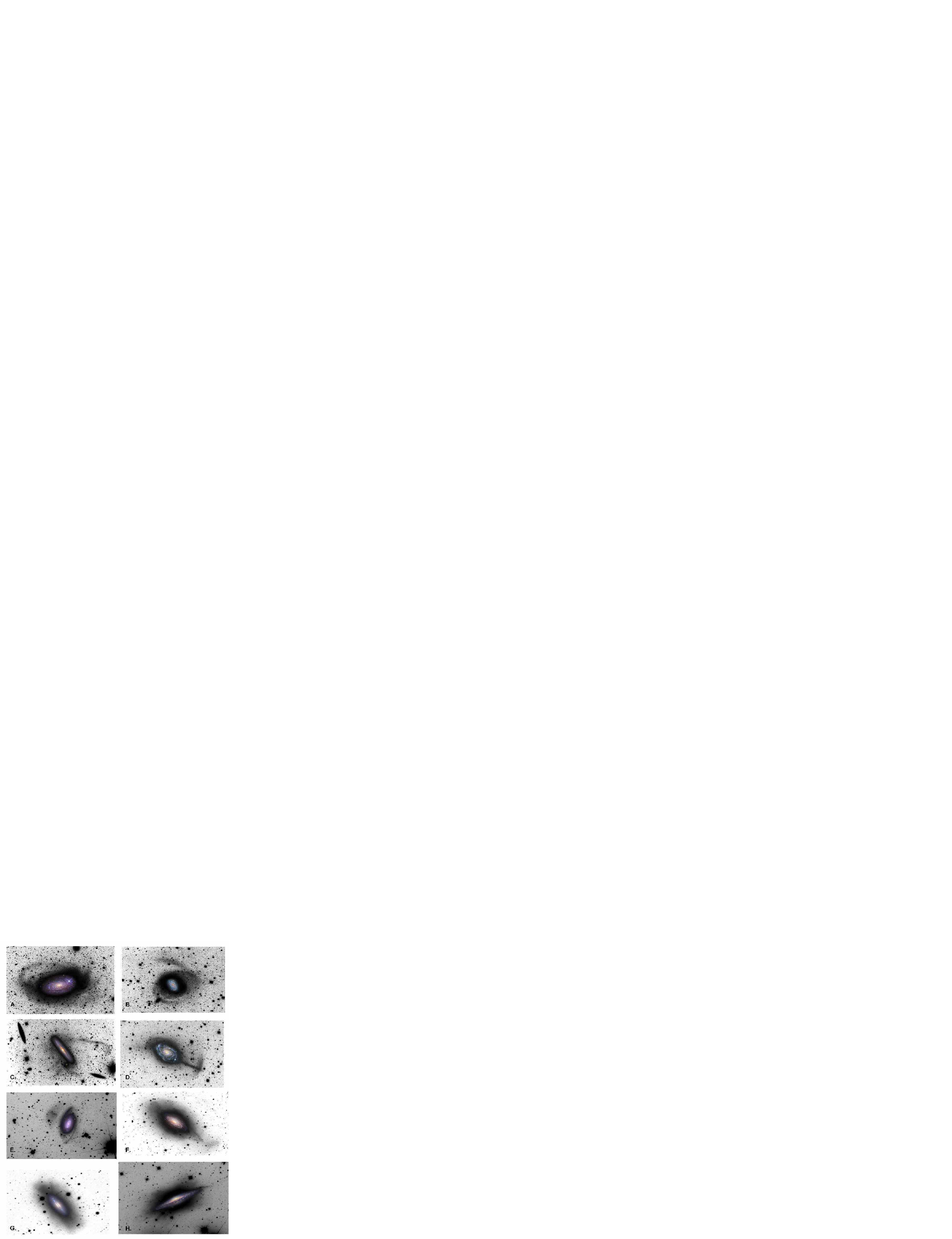}
\caption{\small Luminance filter images of nearby galaxies from our pilot survey (see Sec. 2 for discussion) showing 
large, diffuse light substructures in their outskirts: (a) a possible Sgr-like stream in Messier  63; (b) giant plumes around NGC 1084; 
(c) partial tidally disrupted satellites in NGC 4216; 
(d) an umbrella shaped tidal debris structure in NGC 4651;  
(e) an enormous stellar cloud in NGC 7531; (f) diffuse, large-scale and more coherent features around NGC 3521; (g) a prominent spike and giant wedge-shaped structure seen emanating from NGC 5866 (BBO 0.5-meter); 
  and (h) a strange inner halo in NGC 1055,
  sprinkled with several spikes of debris (RdS 0.5-meter). 
  Each panel displays a (linear) super-stretched 
   contrast  version of the total image. 
   A color
  inset of the disk of each galaxy (obtained from data from
  the same
  telescope as the luminance images) has been over plotted for reference
  purposes. In addition, some of the original  images were also cropped to
  better show the most interesting regions around each target.\label{fig1}}
\end{figure}

\begin{landscape}
\begin{table}
\small
\caption{Observatories and other facilities}
\label{obs}
\begin{tabular}{l l l c c c c }
\hline
\hline
Observatory & Location  & Telescope & focal ratio & CCD & Scale
($\arcsec$/pix) & FOV (arcmin) \\ 
\hline
\noalign{\smallskip}
Black Bird (BBO) &  New Mexico, USA & RCO 0.508-m & f/8.1&  STL-11000
& 0.46 & 20.4 $\times$ 30.6 \\
Rancho del Sol (RdS) &  California, USA & RCO 0.508-m & f/8.1 & Alta
KAF09000 & 0.58 & 29.3 $\times$ 29.3 \\ 
Moorook (Mrk) & South Australia & RC0 0.368-m &f/9 & Alta 16803 & 0.56 &
38.2 $\times$ 38.2 \\
New Mexico Skies (NMS) &  New Mexico, USA & APS 0.160-m &
f/7 & STL-11000 & 1.66 & 73.7 $\times$ 110.6 \\
\hline
\end{tabular}
\end{table}
\end{landscape}

\begin{table}
\centering
\label{obs}
\begin{tabular}{l l c  c  c}
\hline
\hline
Target & Facility  &  Exp. time (L) & FOV (arcmin) & Features \\ 
\hline
\noalign{\smallskip}
NGC 5055 & NMS 0.16-m &  790  & 37 $\times$ 27  & A \\
NGC 1084 & BBO 0.51-m & 825   & 15 $\times$ 10  & GP \\ 
NGC 4216 & RdS 0.51-m & 1105 & 28 $\times$ 27   & PD \\
NGC 4651 & BBO 0.51-m &  585 & 14 $\times$ 9    & U \\
NGC 7531 & Mrk 0.37-m &  600 & 14 $\times$ 10   & Cl \\
NGC 3521 & BBO 0.51-m & 495  & 21 $\times$ 16   & Mx, U? \\ 
NGC 5866 & BBO 0.51-m & 315  & 9 $\times$ 7     & Sp, W \\
NGC 1055 & RdS 0.51-m & 400 & 13 $\times$ 9      & Mx, Sp \\
\hline
\end{tabular}
\caption {Observational data for galaxies displayed in Fig 1. {\bf Column 1}: NGC number of the target galaxy; {\bf
    Column 2}: Observatory name and telescope aperture (see Table 1); {\bf Column 3}:
total exposure time in minutes of the luminance image plotted in Fig. 1;{\bf
  Column 3}: Field of view in arc minutes of the zoomed image displayed in
Fig. 1; {\bf Column 5}: morphology
of the tidal
features (see Sec.~3) recognizable in the outskirts of the galaxy (A = arcs; GP = giant
plumes; PD= partially disrupted satellite; 
U = umbrella; W = wedge; Cl = cloud; Sp = spike;
Mx = mixed-type).} 
\end{table}

\newpage

\begin{figure}[!]   
\epsscale{.95}
\plotone{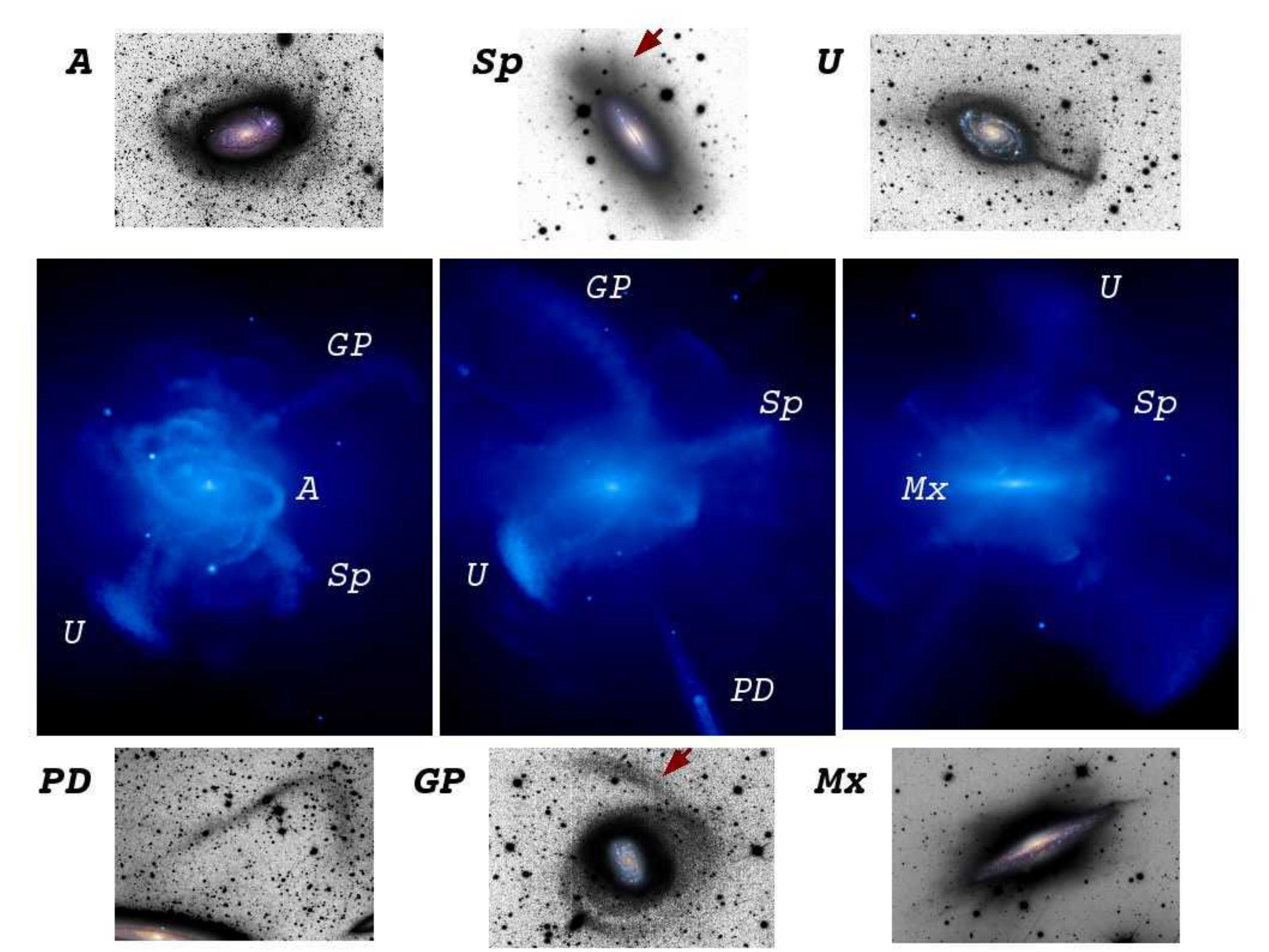}
\caption{{\small An illustrative comparison of some of the different features detected in our 
pilot survey to the surviving structures visible in cosmological simulations of 
Milky Way-like galaxies (see Sec.~4).
The three
central panels provide an external perspective realized through  a simulation (from left to
right, halo models numbered 17, 15  and 20 from Johnston et al. 2008) within the
hierarchical framework and show luminous streams resulting from tidally disrupted
satellites. Each {\it snapshot} is 300 kpc on a side. 
The tidal features labeled
in the snapshots identify structures similar to those observed in our data:
{\it A}  great circles features (Messier 63); 
{\it Sp} spikes (NGC 5866); 
{\it U} umbrella shaped structures (NGC 4651);
{\it PD} partially disrupted satellites (NGC 4216); 
{\it GP} giant plumes (NGC 1084); 
{\it Mx} possibly mixed type streams (NGC 1055).\label{fig2}}}
\end{figure}

\end{document}